\documentclass[%
 reprint,
 amsmath,amssymb,
 aps,
]{revtex4-1}

\usepackage{graphicx}% Include figure files
\usepackage{dcolumn}% Align table columns on decimal point
\usepackage{bm}% bold math
\usepackage{float}
\usepackage{listings} % for displaying code
\usepackage{verbatim}
\begin{document}

\preprint{APS/123-QED}

\title{Compressed Sensing for STM imaging of defects and disorder}% 
\thanks{This manuscript has been authored by UT-Battelle, LLC under Contract No. DE-AC05-00OR22725 with the U.S. Department of Energy. The United States Government retains and the publisher, by accepting the article for publication, acknowledges that the United States Government retains a non-exclusive, paid-up, irrevocable, world-wide license to publish or reproduce the published form of this manuscript, or allow others to do so, for United States Government purposes.  The Department of Energy will provide public access to these results of federally sponsored research in accordance with the DOE Public Access Plan (http://energy.gov/downloads/doe-public-access-plan).}%

\author{Brian E. Lerner}
\author{Anayeli Flores-Garibay}%
\author{Benjamin J. Lawrie}
\author{Petro Maksymovych}\email{maksymovychp@ornl.gov}

\affiliation{%
 Oak Ridge National Laboratory, 1 Bethel Valley Rd, Oak Ridge, TN 37831
}%

\date{\today}% It is always \today, today,

\begin{abstract}
Compressed sensing (CS) is a valuable technique for reconstructing measurements in numerous domains. CS has not yet gained widespread adoption in scanning tunneling microscopy (STM), despite potentially offering the advantages of lower acquisition time and enhanced tolerance to noise. Here we applied a simple CS framework, using a weighted iterative thresholding algorithm for CS reconstruction, to  representative high-resolution STM images of superconducting surfaces and adsorbed molecules. We calculated reconstruction diagrams for a range of scanning patterns, sampling densities, and noise intensities, evaluating reconstruction quality for the whole image and chosen defects. Overall we find that typical STM images can be satisfactorily reconstructed down to  30\% sampling - already a strong improvement. We furthermore outline limitations of this method, such as sampling pattern artifacts, which become particularly pronounced for images with intrinsic long-range disorder, and propose ways to mitigate some of them. Finally we investigate compressibility of STM images as a measure of intrinsic noise in the image and a precursor to CS reconstruction, enabling a priori estimation of the effectiveness of CS reconstruction with minimal computational cost.  
\end{abstract}

\maketitle

%\tableofcontents

% THRUSTS
% Atomically resolved data
% Phase diagrams
% Framework / Admittance criterion 

\section{Introduction}

Scanning tunneling microscopy (STM) and spectroscopy (STS) have become indispensable techniques for electronic, structural and magnetic characterization of surfaces with atomic resolution. STM has enabled investigations of broken symmetry and vortex interactions in superconductors \cite{Hoffman2011, Fischer2007}, enabled the band structure mapping of quantum materials \cite{Oppliger2020}, and was used for the first observations of spatial LDOS modulations \cite{Hasegawa1993, Crommie1993}. 

However, small tunneling currents limit the rate of current measurement to the millisecond timescale, so that STM measurements are characterized by comparatively long measurement times \cite{Oppliger2020}. This limitation becomes apparent in experiments that seek to probe extended surface areas, seek rare events such as low density defects, and want to strike a balance between high-resolution measurement in real space and energy resolution. In such cases, the ability to accurately reconstruct the underlying periodic and defect structure of nanoscale samples with reduced measurement time is highly desirable.

Compressed sensing (CS) shows potential for meeting this demand. CS is based on the notion that if a basis set can be found where the signal is sparse (and as a corollary the signal is compressible in that basis), accurate reconstruction is possible using fewer measurements than required by the Shannon-Nyquist Sampling Theorem. CS has been successfully employed for diverse applications including radio interferometry \cite{Honma2014}, nuclear magnetic resonance of protein structure \cite{Kazimierczuk2011, Holland2011}, recovery of correlations of entangled photon pairs \cite{Simmerman2020,lawrie2013toward}, medical imaging \cite{Lustig2007, Lustig2008} and many more. 

 An image is compressible by virtue of its sparsity in a transform domain. Most images in the natural world have a sparse frequency or wavelet representation, including those generated by scanning microscopies. Indeed, CS has been successfully implemented in scanning electron \cite{He2009}, atomic force \cite{Oxvig2017}, and piezoresponse force microscopy \cite{Kelley2020}, 
%as well as QPI patterns from 
and quasiparticle interference imaging by STS \cite{Oppliger2020,Nakanishi-Ohno2016}. However, a detailed understanding of the potential of CS for STM has yet to be developed, particularly with respect to imaging defects and disorder.

In this paper, we explore the parameter space of a simple CS framework in the context of representative STM images from surfaces of superconductors and single molecule layers (introduced in section  \ref{sec:experimental-data}). Our specific focus is to emphasize the quality of reconstruction around defects and as a function of added noise. In sections \ref{sec:CS-basics} and \ref{sec:CS-framework}, the basic methodology of CS is laid out, and the framework is described.  Using a soft weighted iterative thresholding (SWIT) algorithm  of practical computational complexity, we performed reconstructions across variable noise perturbation intensities and sampling densities. These reconstructions are evaluated for structural similarity index measure (SSIM) and mean squared error (MSE) and are used to calculate reconstruction diagrams in section \ref{sec:results}. Our results reveal that accurate reconstruction can be obtained at sampling densities as low as 20-30\% for images with both point and extended nanoscale defects - i.e. with almost 5-fold compression. We also note artifacts arising in the reconstructions, and detail ways of mitigating these deviations through proper algorithm configuration. To effectively apply CS in practice, it is very helpful to understand what types of images can be effectively reconstructed. In \ref{sec:results}, we also characterize our images using compressibility, finding compressibility to be an effective measure of noise in the STM images, and a necessary, albeit not sufficient, criterion for effective CS reconstruction. %Since CS is an extensible technique, in section \ref{sec:improvements}, we discuss ways to augment the provided framework through alternative choices of algorithm, basis transform, and sampling pattern.

% talk about "functional complexity" (i.e

\section{Experimental Data} \label{sec:experimental-data}
\begin{figure}[t]
    \centering
    \includegraphics[width=\columnwidth]{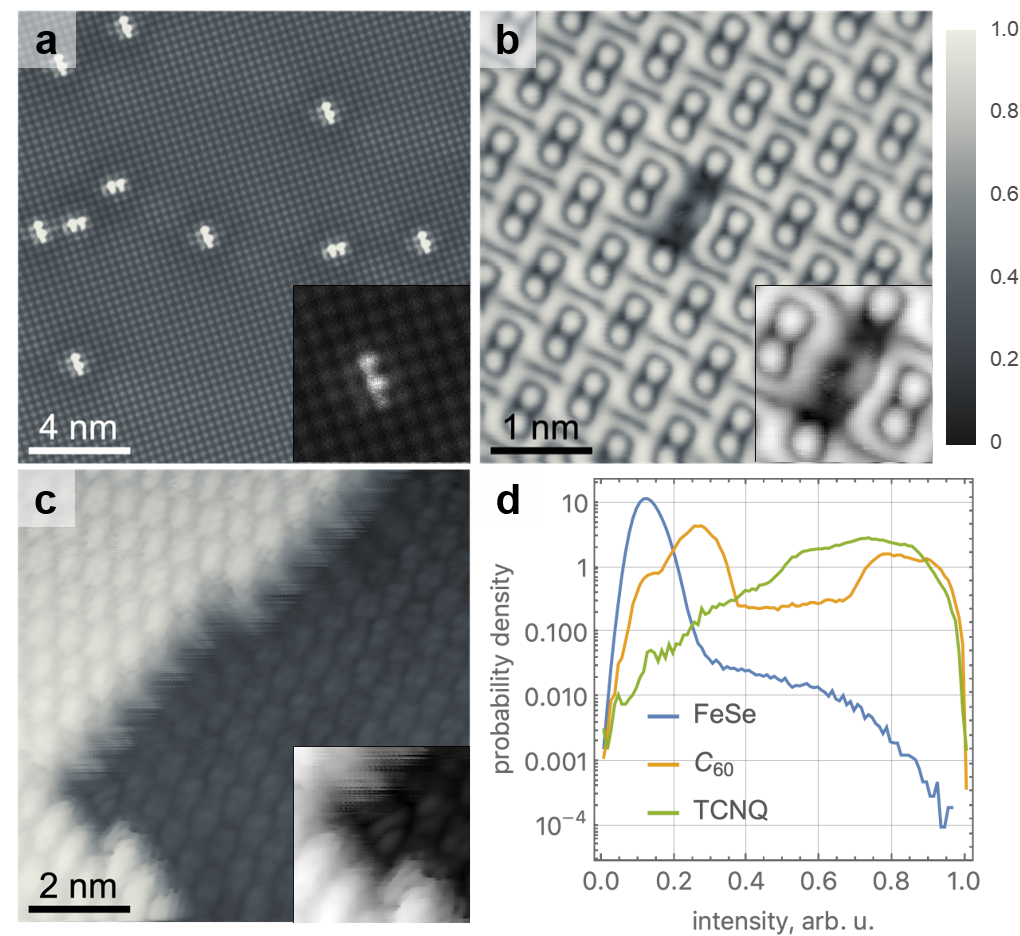}
    \caption{STM images of FeSe (a), TCNQ (b), and C$_{60}$ (c), with representative defects magnified in each inset. (d) The distribution of normalized constant-current STM height for each image.}
    \label{fig:Fig1}
\end{figure}

We applied CS to representative STM images of a cleaved 100-surface of FeSe superconductor with Se vacancy defects \cite{Huang2016} (Fig. \ref{fig:Fig1}a) and two kinds of adsorbed molecular layers - C$_{60}$ on Ag(111) (Fig. \ref{fig:Fig1}c) and TCNQ (tetracyanoquinodimethane) on graphite (Fig. \ref{fig:Fig1}b). Each of the sample images have a different size, lattice structure, and point or extended defect. Moreover, as seen in Fig. \ref{fig:Fig1}d, the images represent three kinds of intensity distribution, centered on low values corresponding to the atomic lattice in the case of FeSe, a broader and more uniform distribution in the case of TCNQ and a distinctly bimodal distribution for C$_{60}$, owing to a single atomic step of the underlying substrate.

\section{CS Basics} \label{sec:CS-basics}

%% [elaborate idea of sparsity]
Sparsity regularization is a common approach to impose constraints on undefined optimization problems \cite{Claerbout1973}, which gave rise to CS methodology in the mid-2000s \cite{Donoho2006, Candes2006}.  CS is designed to reconstruct a signal $x \in \mathbb{R}^{n \times 1}$  from samples $y \in \mathbb{R}^{m \times 1}$, where typically $m \ll  n$. Successful reconstruction is possible when $x$ has a sparse representation $\alpha \in \mathbb{R}^{n \times 1}$, i.e. in some basis the number of significant coefficients $k$ in  $\alpha$ is small compared to $n$. The CS algorithm computes  $\alpha$. Once obtained, $x$ is recovered using the basis transform $\Psi \in \mathbb{R}^{n \times n}$:
\begin{align} \label{eq:x-psi-alpha}
x = \Psi \alpha
\end{align}
The sampling process has a matrix representation $\Phi \in \mathbb{R}^{m \times n}$ constructed by stacking each measurement vector:
\begin{align} \label{eq:phi-x-y}
\Phi x = y
\end{align}
Substituting eq. \ref{eq:x-psi-alpha} for $x$ in eq. \ref{eq:phi-x-y} and setting $A=\Phi\Psi$ we have:
\begin{align} \label{eq:A-alpha-y}
A \alpha = y
\end{align}
 CS provides a solution $\alpha$ for this undetermined system of equations by minimizing the sparsity of $\alpha$ under the constraints of eq. \ref{eq:A-alpha-y}, expressed as:
\begin{align} \label{eq:l0-minimization}
    min \| \alpha \|_{\ell_0} \quad \text{s.t.} \quad A\alpha = y
\end{align}
While this provides an exact solution, $\ell_0$ minimization is a combinatorial optimization problem that is computationally expensive, and intractably so for large signals \cite{Candes2006}. Fortunately, the $\ell_1$ norm can be substituted to convert the problem into one of convex optimization, where for most inputs, $\alpha$ is recovered exactly \cite{Candes2006}. 

% Real signals almost always involve some form of measurement noise. This can be accounted for by introducing a noisy vector $ e \in \mathbb{R}^{m \times 1}$ into eq. \ref{eq:A-alpha-y}:
% \begin{equation} \label{eq:A-alpha-e-y}
% A\alpha + e = y 
% \end{equation}

% % need to better understand noisy model before including
% \begin{comment}
% In the case of such a noisy model, a sparse solution is sought so that:
% \begin{equation} 
% \|y-A\alpha \|\leq \epsilon 
% \end{equation}
% where $\epsilon$ is a bound on the amplitude of the noise.
% \end{comment}
% CS requirements
% There are two essential requirements for successful CS application: 1) sparsity, and 2) incoherence. As described above, CS rests on the notion that the signal of interest is sparse in some domain [build up]. As for incoherence, this denotes that ...[build up].
% % robustness to noise, refer to Candes/wakin

% Different implementations are plentiful, ranging widely in complexity. A brief survey of more advanced methods is given in section \ref{improvements}.

\section{Framework} \label{sec:CS-framework}

The CS framework can utilize a variety of 1) sampling matrices $\Phi$, 2) transform matrices $\Psi$, and 3) optimization algorithms.
%When choosing $\Phi$ and $\Psi$ it is paramount to ensure that they meet the CS requirements outlined in \ref{sec:CS-basics}. 
$\Psi$ should necessarily be chosen to ensure sparsity in the transform domain, but it should also be incoherent with $\Phi$. The algorithm minimizes the sparsity in $\alpha$ while remaining correlated to the measurements $y$ (eq. \ref{eq:A-alpha-y}). In our reconstructions, we use Lissajous and rotated line trajectories for sampling patterns, the discrete cosine transform (DCT), and a SWIT algorithm. The elements of this framework, with special regard to their applicability for STM, are discussed in the following.

\subsection{Transform Matrix}

STM images often exhibit a large amount of order and are generally smooth (i.e. differentiable in the absence of noise). As a result, the images lend themselves to sparsity in the DCT basis. The DCT transform matrix also has the advantage of being maximally incoherent with point sampling matrices \cite{Candes2008}, and has a fast matrix implementation \cite{Arildsen2015}. This transform has been utilized in previous applications of CS \cite{Romberg2008, Jensen2013, Anderson2013}, and has historically been used for JPEG compression \cite{Wallace1992}. The discrete wavelet transform (DWT) is another commonly used dictionary in compressed sensing, thought it works most efficiently with dense sampling matrices with random entries like those used for single-pixel imaging and is less incoherent than DCT for point sampling matrices \cite{Arildsen2015}.

\begin{figure}%[b!]
    \centering
    \includegraphics[width=\columnwidth]{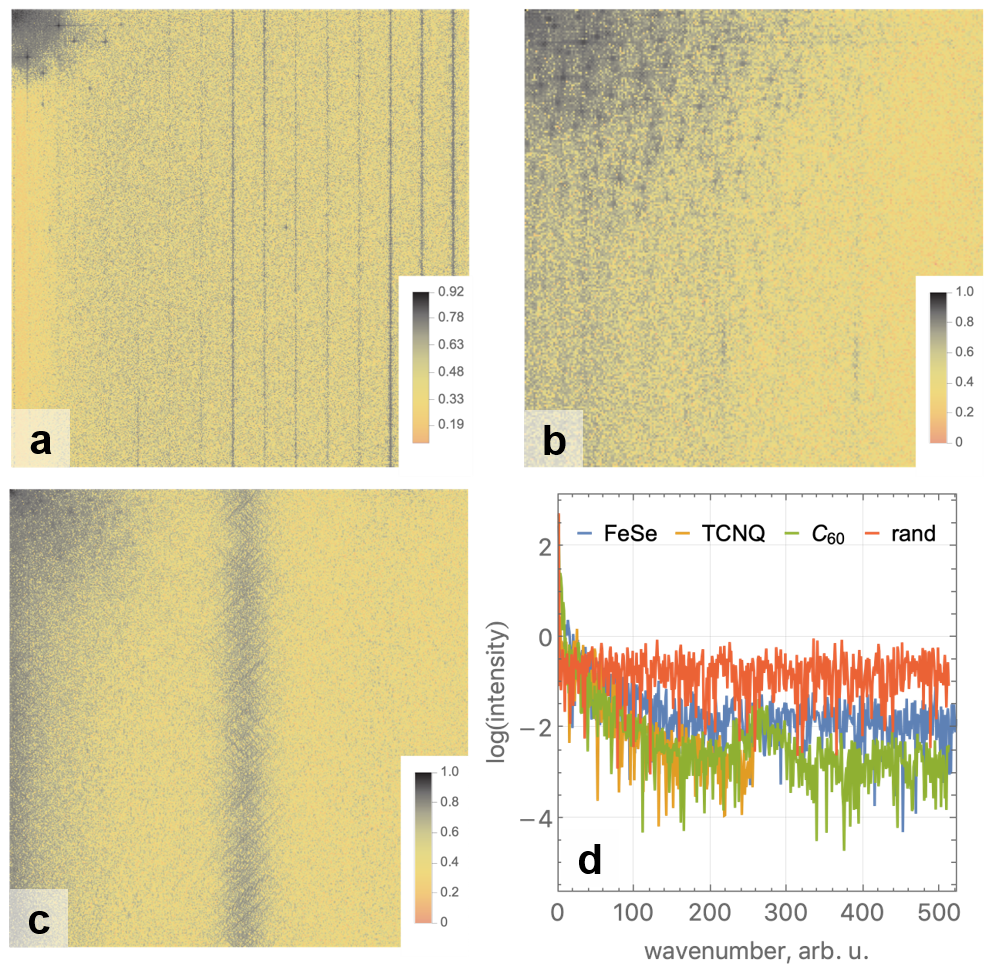}
    \caption{DCTs of (a) FeSe, (b) TCNQ, and (c) C$_{60}$. (d) The intensity of the diagonal coefficients for each DCT, as well as the DCT of an array of random Gaussian noise, which demonstrate varying sparsity levels.}
    \label{fig:Fig2}
\end{figure}

%In JPEG, the DCT is applied to sections of the image but has been previously shown to work well when applied to the whole image \cite{Arildsen2015}. The same is done in our study. 

%  Different dictionaries can be used for a sparse representation in some basis. Some are fixed, general purpose and orthogonal \cite{Arildsen2015}. We believe some of these dictionaries can be used to further optimize our transform matrix beyond the DCT. 

\subsection{Sampling Matrix}

\begin{figure}%[t!]
    \centering
    \includegraphics[width=\columnwidth]{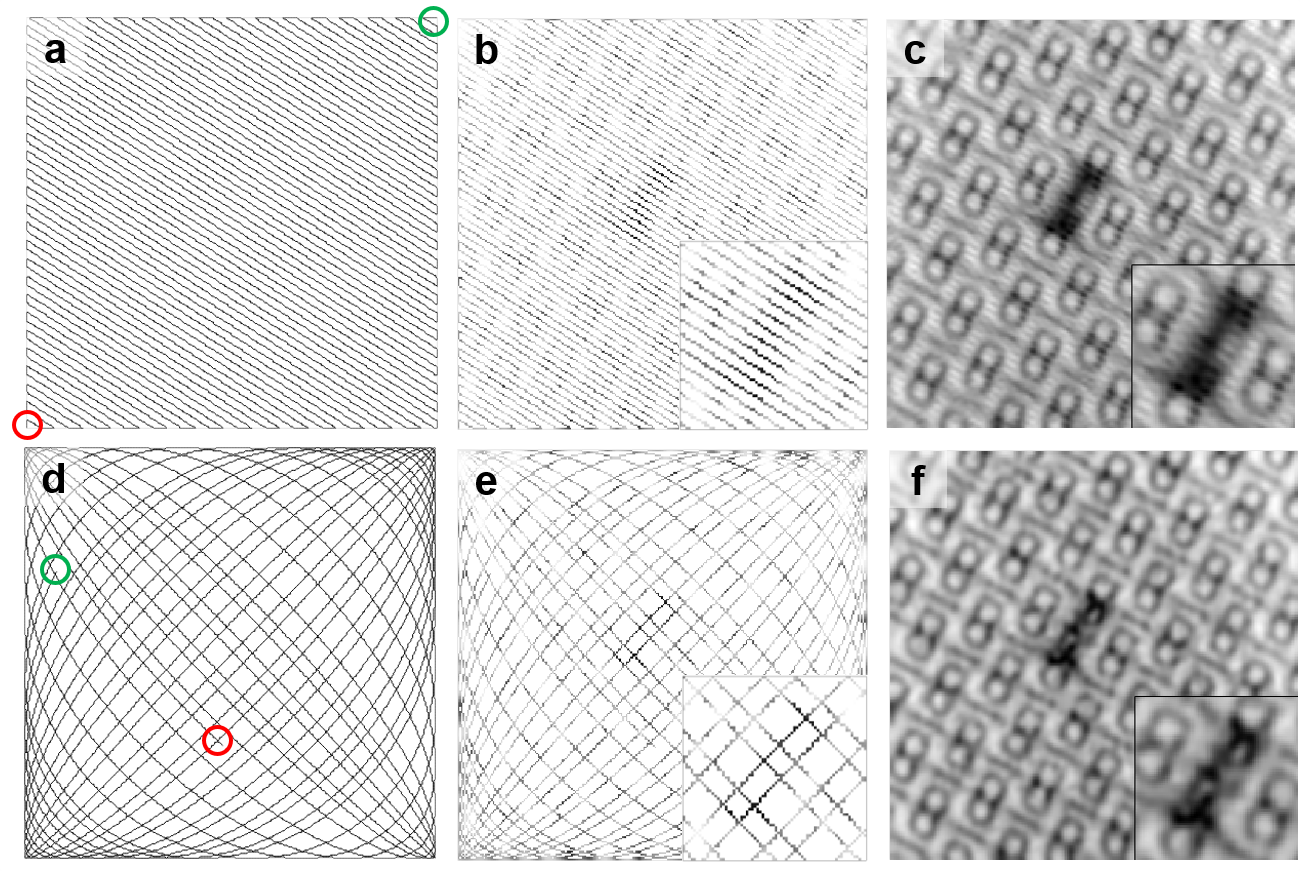}
    \caption{The path of the rotated line pattern is shown in (a), with simulated start and end points denoted by green and red circles. Despite sparse sampling of the image (b), decent reconstruction is achieved (c). The same process is also shown for Lissajous (d-f). Reconstructions in this figure performed for 20\% sampling density and 100 iterations. }
    \label{fig:Fig-Patterns}
\end{figure}

When scanning a surface, it is conventional to use a raster scan, where the probe traverses the sample in a series of alternating lines, resulting in an evenly sampled grid. The speed of the probe and the sampling frequency are set based on the demands of the experiment. While the design of the sampling matrix $\Phi$ in other CS applications is often flexible (programmable with a spatial light modulator for optical CS applications, for instance), we are constrained to sampling along the continuous path of the probe. Here, since we are concerned with the algorithmic aspects of the reconstruction, we chose to use pre-existing STM images and resample them with smooth Lissajous (Fig. \ref{fig:Fig-Patterns}d) and rotated line (Fig. \ref{fig:Fig-Patterns}a) patterns which make the methods more compatible with fast scanning. The sampling can furthermore be randomized along the sampling path, but we have not seen a significant impact from such randomization.

\subsection{Optimization Algorithm} % is optimization always true for recon algorithm?
There are a variety of reconstruction algorithms that have already been explored for other CS applications. In the convex optimization class, the $\ell_0$ norm is replaced by the $\ell_1$ norm. Greedy pursuit algorithms use an iterative approach where locally optimal decisions are made in each iteration. Iterative thresholding \cite{Herrity2006} is a type of greedy pursuit algorithm that has relatively low computational complexity and is robust to noise. %[CITE]. 
Due to these benefits, we employed a SWIT algorithm as successfully demonstrated in \cite{Oxvig2017}. The algorithm works as follows:
\vspace{5mm}
\begin{lstlisting}[language=Python, mathescape=true, aboveskip=4pt,belowskip=0pt, label={lst:SWIT}] 
$\alpha = 0$
$r = y$
for i in I:
    $c = A^T r$
    $\alpha = \eta_t^{ws}(\alpha + \kappa \cdot c )$
    $r = y - A \alpha$
    if $\| r \|_{\ell_2}  < \epsilon \| y \|_{\ell_2}$:
        break
\end{lstlisting}
\vspace{2mm}

Initialization to $\alpha = 0$ can be changed to an educated guess and the stopping condition can be arbitrarily chosen, while the step size $\kappa$ ensures convergence. % [CITE]. 
The soft weighted thresholding function $\eta_t^{ws}$ is implemented as: %[CONFIRM]:
\begin{align} \label{eq:threshold-function}
    \eta_t^{ws} &= \frac{1}{w}sgn(x)(|wx| -t), |wx| -t > 0 \\
    &= 0, |wx| -t \leq  0
\end{align}

\noindent The method for calculating the threshold $t$ is customizable. Here, we set a fixed value on the number of nonzero coefficients while initializing the algorithm. In each iteration, the coefficients are weighted as described above, $t$ is adjusted to maintain the specified sparsity, and coefficients below $t$ are zeroed. By tuning the weights to model expected DCT dispersion, weighted iterative thresholding algorithms tend to outperform their non-weighted counterparts \cite{Oxvig2017}. Each of the reconstructions constituting the reconstruction diagrams ran for 100 iterations due to computational considerations, though in our experiment we found that reconstruction tends to improve up to around 300 iterations--and sometimes many more--before plateauing.  %The weights serve only to determine which coefficients are set to 0, while having no effect on the rest. 

\begin{figure}
    \centering
    \includegraphics[width=\columnwidth]{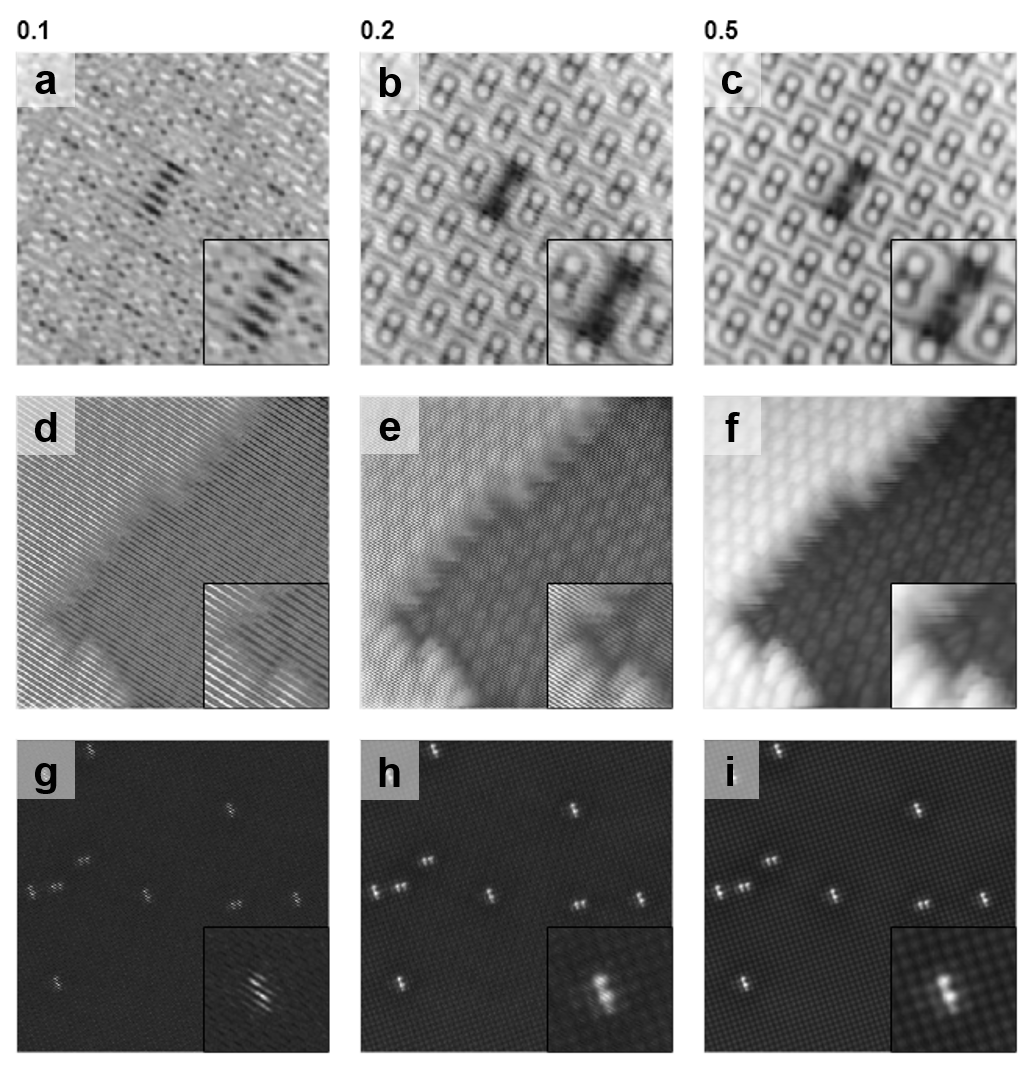}
    \caption{Reconstructed images for ten, five, and two-fold undersampling for TCNQ (a-c), C$_{60}$ (d-f), and FeSe (g-i), with magnified defects in insets. All reconstructions performed for 100 iterations using the rotated line sampling pattern.}
    \label{fig:Fig3}
\end{figure}
\subsection{Quality Assessment} \label{sec:methods}

To understand the bounds of reconstruction, we evaluated the SWIT algorithm while systematically varying the noise intensity $\delta$ and sampling density $\rho$. 
%The original data constituting the selected images was obtained in a raster pattern [CONFIRM] and sampling was performed post-measurement. 
% While this does not facilitate an exact number on measurement timing, the relationship between number of samples and measurement duration is clear.
% cite Arildsen2015 about coolwarm being better)
While iterative thresholding algorithms are noted for being noise-robust \cite{qu2010iterative}, little investigation has been carried out to confirm this for reconstruction of STM images. In order to test this, we generated $1/f$ noise in Python and applied it to pixels along the simulated measurement path so as to mimic varying noise levels during measurement. The noise perturbation scale for each image was normalized to range from 0.1--1 of the highest-peak FWHM in the image's intensity histogram (Fig. \ref{fig:Fig1}d). We used FWHM as a measure of spread due to the approximately Gaussian shape of the distributions. We implemented rotated line and Lissajous sampling patterns across $\rho$ from 0.02--0.5. The patterns used here were generated using \textbf{magni} \cite{Oxvig2014}, a compressed sensing Python package for atomic force microscopy.

For each reconstruction in this $\delta$--$\rho$ parameter space, the quality of the reconstructed image was evaluated for SSIM and MSE. SSIM was calculated using \textbf{scikit-image's} \cite{scikit-image} default implementation, which is adapted from \cite{Wang2004}. The MSE is derived in the standard way,
\begin{align}
    \frac{1}{N}\sum ( \chi - x )^2
\end{align}
where $N$ is the number of pixels, $x$ is the reconstructed image and $\chi$ is the base image. % pros/cons of using MSE
% While the MSE has its drawbacks \cite{Wang2009}%[develop]
% , it is a simple and practical method for understanding the performance of the algorithm in different regimes for particular images, though cross-comparison of different images is not meaningful. 

% if I end up using the same configuration values as Magni, make sure to give credit
%The following inputs determine the path of the sampling pattern: image dimensions and scan length, both in pixels, and the number of samples. The path can be configured to follow the edge [for all patterns?], and pattern-specific options can be set such as the line angle for the uniform rotated line pattern. 

% \subsection{Algorithm configuration}
To perform these reconstructions, we build our CS framework with a DCT transform, due to the benefits espoused in Sec. \ref{sec:CS-framework}, in combination with the noted sampling patterns. We solve eq. \ref{eq:A-alpha-y} using a SWIT algorithm as described in the code block in Sec. \ref{sec:CS-framework}, with $\kappa=0.6$ and a stopping condition that occurs when the ratio of the 2-norm of the residual ($y-Ax$) and the 2-norm of $y$ is less than a tolerance $\epsilon = 0.001$. The weights of the soft thresholding function $\eta_t^{ws}$ used in these reconstructions are adopted from a Gaussian model of DCT structure in \cite{Oxvig2017}, which was used to successfully reconstruct AFM images. 

\begin{figure}[b!]
    \centering
    \includegraphics[width=\columnwidth]{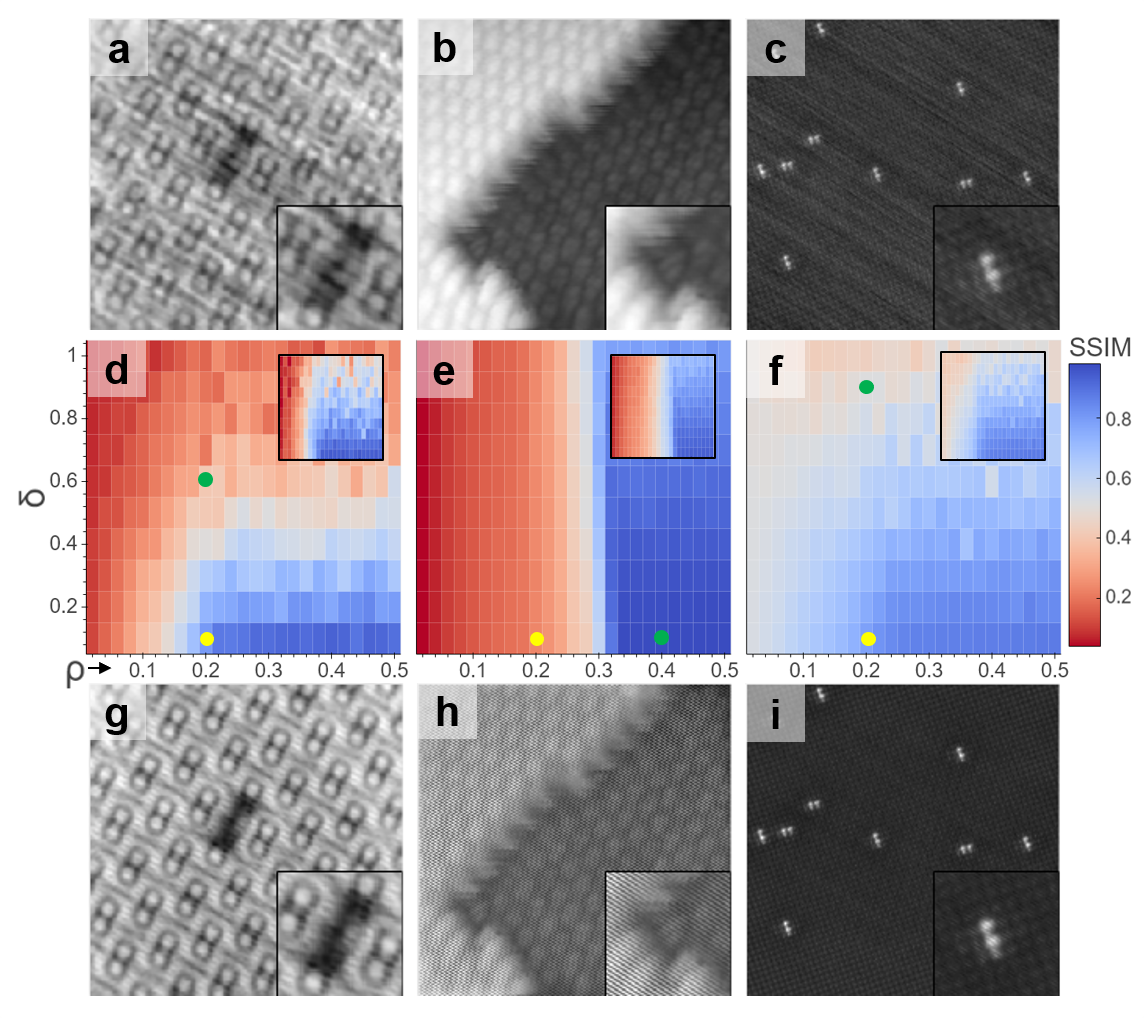}
    \caption{Noise perturbation intensity ($\delta$) vs. sampling density ($\rho$) phase diagrams for reconstructions of TCNQ (a), C$_{60}$ (b) and FeSe (c), with relevant reconstructions shown above and below the diagram for each sample. The parameters used for the reconstructions in the top row are marked by green dots in the respective diagrams; the bottom row parameters are marked by yellow dots.}
    \label{fig:Fig4}
\end{figure}

\section{Results} \label{sec:results}

Our first observation is that CS is generally very good at reconstructing STM images even at a sampling density as low as 20\% of the original image. To ascertain that this conclusion applies not only to spatial order in the images, but also to defect sites, we have identified the latter using state-pace methods for detection of protrusions (using Laplacian of Gaussian filter), and then built local masks of the defects, comparing reconstruction in that local region. As seen in the insets of Fig. \ref{fig:Fig3}, single vacancies in FeSe and extended defects in the TCNQ overlayer (missing molecules) reconstruct well. At 50\% sampling density, the reconstructed defects are indistinguishable from their unsampled counterparts.

%Phase transition diagrams have been used to plot the performance of reconstruction algorithms by varying sampling density and sparsity. % [CITE]. 

The $\delta$--$\rho$ reconstruction diagrams (Fig. \ref{fig:Fig4}) demonstrate the method's robustness to moderate $1/f$ noise. All reconstructions have high SSIM above sampling density $\rho \approx 30\%$ which only begins to degrade at noise perturbations of 0.4 for TCNQ and 0.8 for FeSe. While high-noise distortions are apparent in the reconstructions of TCNQ (Fig. \ref{fig:Fig4}a) and FeSe (Fig. \ref{fig:Fig4}c), the simplicity of FeSe's vacancies and the regularity of its lattice likely lead to smoother SSIM falloff at high noise.  C$_{60}$, in stark contrast, has a wholly noise-independent transition (Fig. \ref{fig:Fig4}e). 
C$_{60}$ also exhibits a sharp transition to higher SSIM at sampling density around 30\%, which exceeds the transition point of the other samples by 10-20\%. Visual examination of the reconstructions (Fig. \ref{fig:Fig4}h) reveals the presence of sampling pattern artifacts at low SSIM which disappear after the transition line. The reasons for this deviation will be discussed below.

% DON'T SPECULATE. AND IF YOU DO, DON"T STOP at ONE-lINER>. I recommend REMOVE this. We will talk about c60 later.
% This is due to the large-scale structure of the step defect cutting across the sample which mitigates noise corruption, and to the relatively small $\delta$ scale used for $C_60$ (Fig. \ref{fig:Fig1}). 

\begin{figure}[b]
    \centering
    \includegraphics[width=\columnwidth]{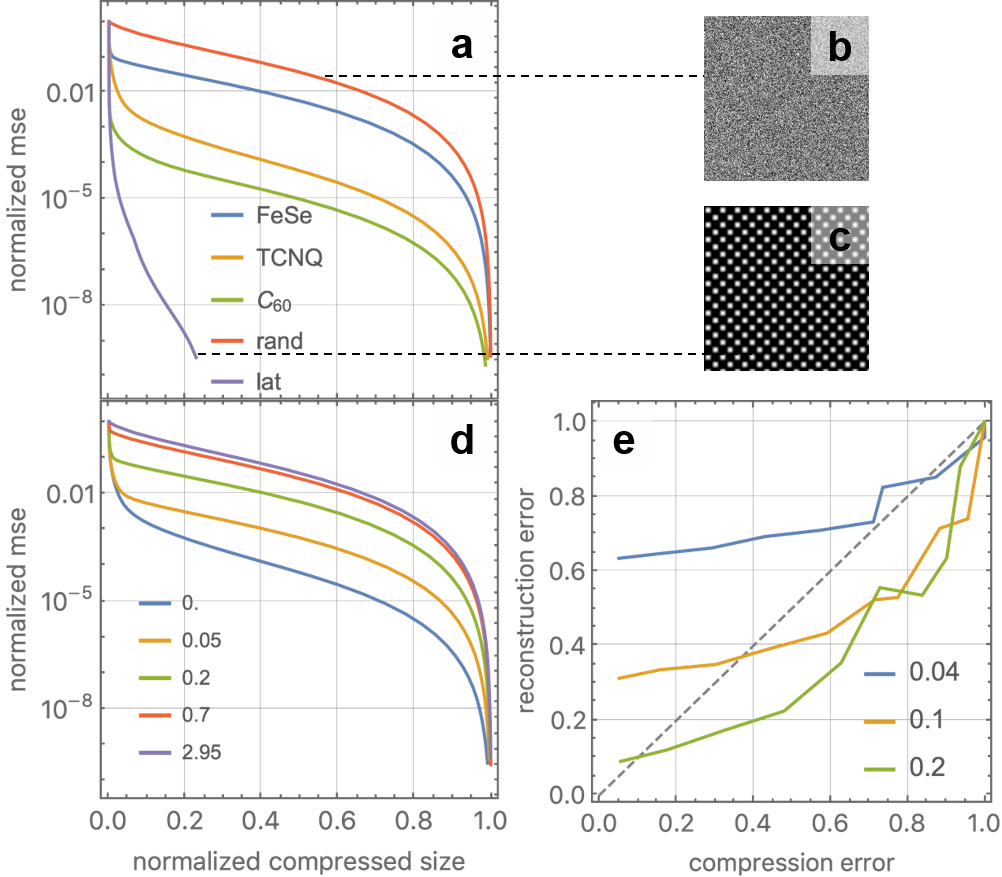}
    \caption{The STM images, along with random Gaussian noise (b) and an ordered lattice (c), were transformed into the DCT basis before being compressed and inverse transformed. The MSE--normalized against the highest value of each curve-- vs. the normalized compressed size, i.e. the compression ratio in the DCT domain, is shown for each image in (a). This procedure is repeated for different levels of Gaussian noise applied to TCNQ pre-transform (d). (e) Compression error vs. CS reconstruction error as a function of noise for varying sampling/compression ratio.}
    \label{fig:Fig5}
\end{figure}
Given that CS is predicated on the principle of compression, we explored the extent to which our CS results correlate to image compressibility for typical STM images as well as simulated arrays, one composed of pseudo-random Gaussian noise (Fig. \ref{fig:Fig5}b) and the other an ordered lattice (Fig. \ref{fig:Fig5}c). We evaluated compressibility by transforming each image and kept a compressed set of the most significant coefficients, setting the rest to 0 before transforming back to real space and evaluating the MSE. The pseudo-random noise image displays the highest error across compression sizes, i.e. it is the most incompressible, while the ordered lattice is most compressible. STM images fall between these two extremes, as seen in Fig. \ref{fig:Fig5}a. Intriguingly, there is a very significant difference between individual images, which actually goes against the trend that may be inferred from the visual inspection of the original data in Fig. \ref{fig:Fig1}. C$_{60}$, not TCNQ or FeSe, is the most compressible image, while FeSe is notably less compressible than either TCNQ or C$_{60}$. 
The difference in compressibility stems from the signal to noise ratio that characterizes these images. To ascertain that this is the case, in Fig. \ref{fig:Fig5}d, we plot compressibility of TCNQ as a function of strength of added noise (measured as a fraction of the largest signal in the image). The compressibility curve very clearly traverses the range observed in Fig. \ref{fig:Fig5}a, eventually becoming equivalent to noise. We note that all these images were all acquired on different days, with different physical tips and different instrument conditions. The ability to ``calibrate" the STM image with compressibility appears to be a valuable measure of the data quality and experimental results.

% \begin{figure}[b]
%     \centering
%     \includegraphics[width=\columnwidth]{img/App_Iterations.png}
%     \caption{For TCNQ reconstructions using a threshold ratio of 5\%, light artifacts appear at 100 iterations (a) that disappear by 200 (b). For C$_{60}$ reconstructions using a threshold ratio of 1\%, heavier artifacts show at 100 iterations (a) that are removed by 300 (c). All reconstructions in this figure were computed with the same DCT weight model as used for the phase diagrams.}
%     \label{fig:Iterations}
% \end{figure}

We now show that the compressibility of an image generally correlates with its CS performance. In Fig. \ref{fig:Fig5}e we plot the normalized CS reconstruction error vs the normalized DCT compression error as a function of noise, for three levels of data compression. For 5-fold compression (20\% sampling), the correlation is reasonably good, which confirms our notion. However, for smaller densities, CS systematically produces higher error  than obtained by DCT compression, which reduces the correlation between the two techniques. We speculate that partly these deviations are due to CS being sensitive to the compatibility of sampling and transform matrices with both each other and the image, as well as the algorithm type and configuration.

\begin{figure}[t]
    \centering
    \includegraphics[width=\columnwidth]{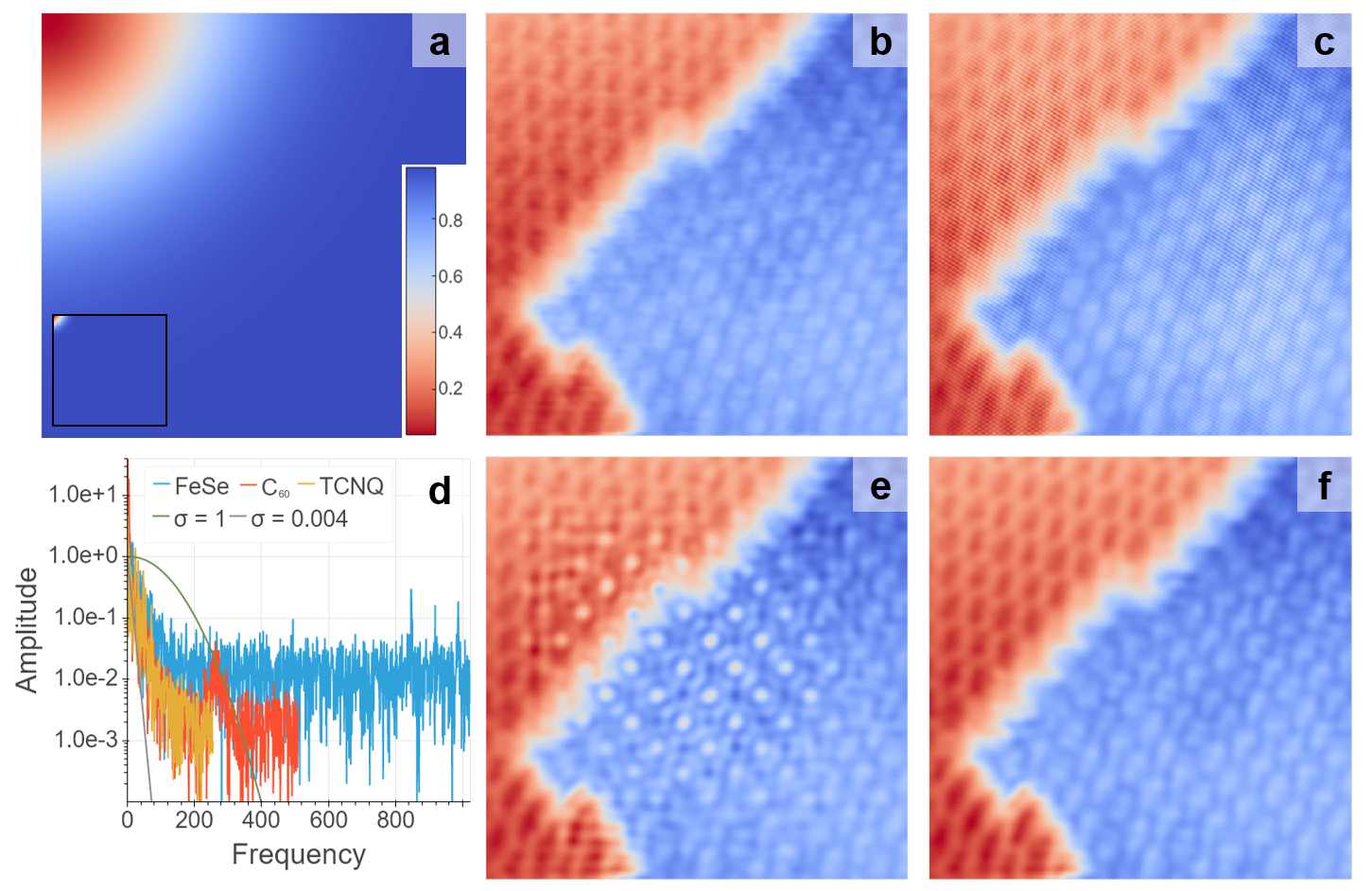}
    \caption{C$_{60}$ reconstructions for Lissajous (b,e) and rotated line (c,f) sampling patterns using a 1\% threshold on the number of non-zero coefficients and 300 iterations. The top reconstructions utilized a wide-variance DCT weight model (a) which was also used for the reconstructions in Fig. \ref{fig:Fig4}. Those on the bottom utilized a model with a severely limited variance; the relevant low-frequency corner of this model is shown in the inset of (a). The diagonal of each model and sample DCT is compared in (d).}
    \label{fig:Fig6}
\end{figure}

A striking disparity, however, appears for C$_{60}$, which is the most compressible of the typical STM images (Fig. \ref{fig:Fig5}a) but requires the highest sampling density to achieve quality reconstruction. Interestingly, past the transition line in both SSIM (Fig. \ref{fig:Fig4}) and MSE (Fig. \ref{fig:MSE_Phase_Diagrams}) phase diagrams.
C$_{60}$ generally has the highest SSIM, followed by TCNQ then FeSe. Resolving this puzzle depends on an understanding of how and when sampling pattern artifacts appear, as their presence is the major cause of $\rho$ dependence in the phase transition. We have found that this brand of artifact can be removed by properly configuring the SWIT algorithm. Small disturbances can be removed by increasing the number of iterations, but more prominent artifacts require increased iterations and/or specialized setup of the threshold function (eq. \ref{eq:threshold-function}).

In each iteration of the SWIT, the threshold function weights each coefficient using a DCT model and based on a specified threshold ratio, keeps a certain number of coefficients while setting the rest to 0. We show that setting the threshold ratio to 1\%, instead of 5\%, running for 300 iterations, and minimizing the variance in the weight model, the artifacts can be removed from C$_{60}$. Reconstruction with the Lissajous pattern was more responsive (Fig. \ref{fig:Fig6}b) to the same DCT-model variance (Fig. \ref{fig:Fig6}a) used for the phase diagrams, though interestingly the rotated line reconstructions improved (Fig. \ref{fig:Fig6}f) only with severely minimized variance (Fig. \ref{fig:Fig6}a inset). 

% Although Fig 7 is 100 iterations, blue SSIM remove defects enough to where iterations will take hold.
% Iterations don't appear for TCNQ because already below to low threshold
% SSIM is not as effective at describing large-scale strcuture

\begin{figure}[b]
    \centering
    \includegraphics[width=\columnwidth]{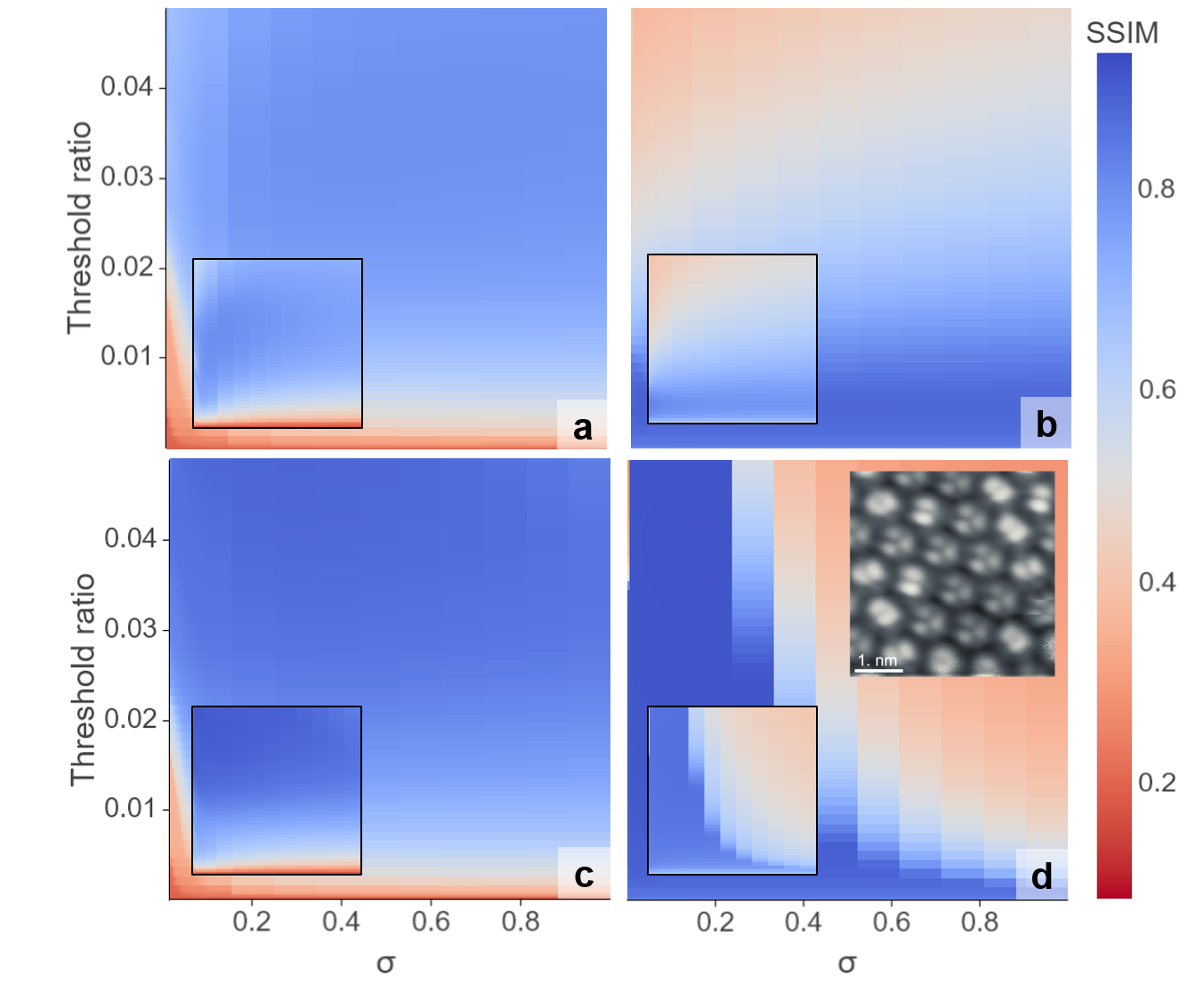}
    \caption{SSIM evaluated for reconstructions of TCNQ (a,c) and C$_{60}$ (b,d) across varying levels of $\sigma$ (the width of the variance in the DCT weight model) and threshold ratio (the relative number of of nonzero coefficients used by the optimization algorithm). The top and bottom rows respectively correspond to reconstructions performed using Lissajous and rotated line sampling patterns. All reconstructions performed with sampling density $\rho = 0.2$.}
    \label{fig:Fig7}
\end{figure}

\begin{figure}[t]
    \centering
    \includegraphics[width=\columnwidth]{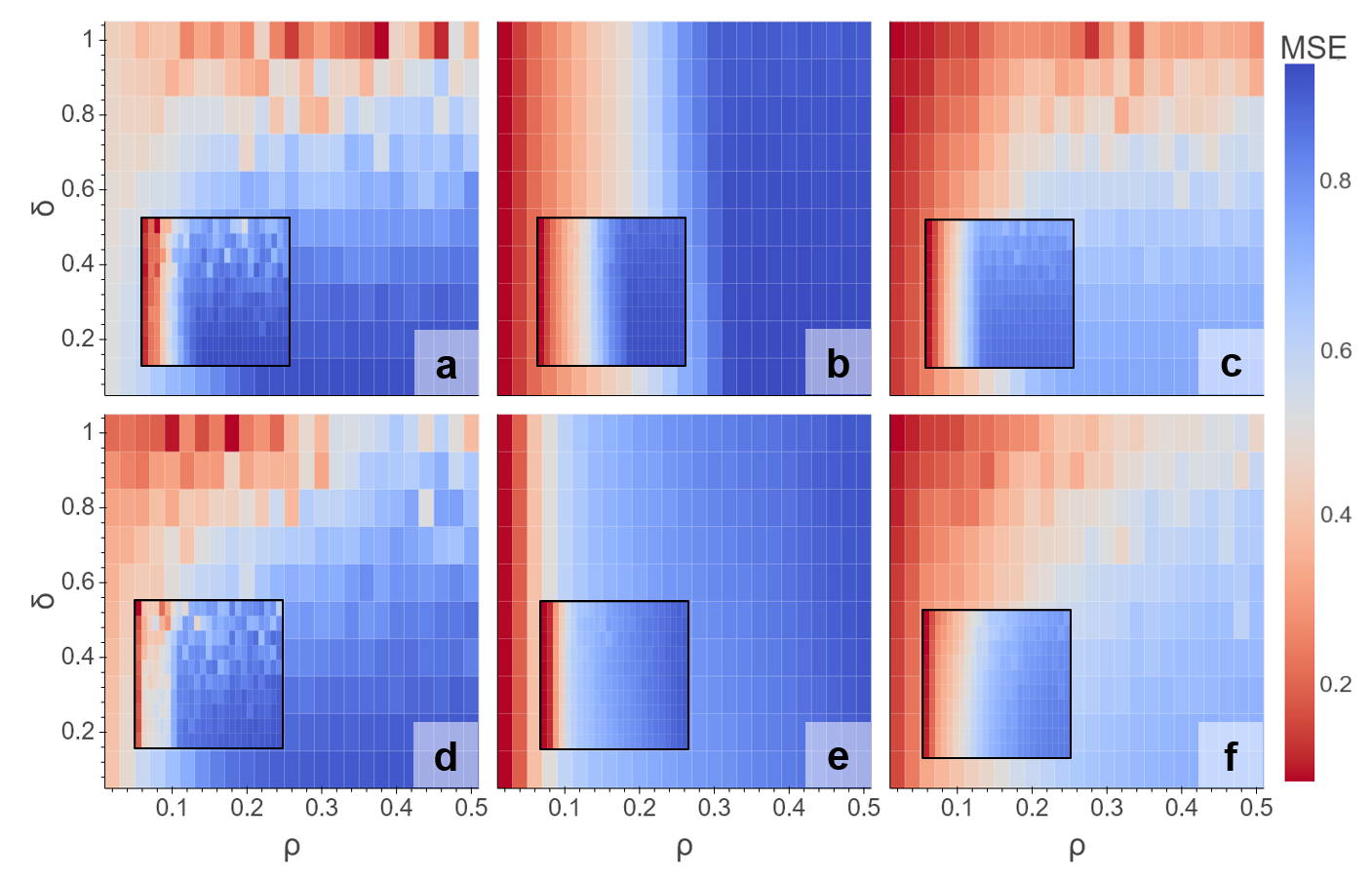}
    \caption{Noise perturbation intensity ($\delta$) vs. sampling density ($\rho$) MSE phase diagrams for reconstructions of TCNQ (a,d), C$_{60}$ (b,e) and FeSe (c,f) for rotated line (top row) and Lissajous (bottom row) sampling patterns with defect phase diagrams in the insets. All phase diagrams have been normalized to their respective maximum MSE.}
    \label{fig:MSE_Phase_Diagrams}
\end{figure}

To determine the ideal thresholding function parameters, we evaluated C$_{60}$ and TCNQ for SSIM across a range of threshold ratios and variances (Fig. \ref{fig:Fig7}). We see that SSIM falls off for TCNQ at low threshold ratios for all variances $\sigma$, and in the limit of low $\sigma$ and threshold ratio-- a trend consistent for both sampling patterns. This behavior is expected as reducing threshold ratio and decreasing $\sigma$ are both tantamount to applying a low-pass filter. Surprisingly, the filtering at low $\sigma$ and threshold ratio produces distinctly higher SSIM for the defect compared to the global image, though visual inspection revels intense lattice warping. The defect diagrams for both samples show higher SSIM for rotated line than Lissajous, a difference especially stark for C$_{60}$. In contrast to TCNQ, which has similar trends in performance for both patterns, C$_{60}$ is quite different. For Lissajous, the SSIM falls off at at threshold ratios around 20\% independently of $\sigma$. Rotated line maintains high SSIM  across low $\sigma$ for all thresholds, though a transition line develops with increasing $\sigma$ that exponentially falls to very low threshold ratios. At low threshold ratio, C$_{60}$ is seemingly immune from SSIM degradation, though the defect diagram has a slight dip at very low threshold. Visual inspection of reconstructions in this regime reveals heavy and unsatisfactory smoothing which retains a semblance of the step defect and an accordingly high SSIM. For all samples and patterns in Fig. \ref{fig:Fig7} though, overlapping high-SSIM regions across global and defect diagrams reveal an optimal parameter space for defect-lattice reconstruction and provide a proof-of-principle for effectively tuning the thresholding function parameters.

To better understand C$_{60}$'s sensitivity to sampling pattern, we refer back to its DCT (Fig. \ref{fig:Fig2}). Each DCT coefficient distribution features a cluster of high-magnitude coefficients in the upper left-hand corner, i.e. for low frequencies. It is important to note that the spread is also dependent on the image dimensions that dictate the full extent of the DCT frequency range. TCNQ and FeSe exhibit denser low-frequency clusters and a scattering of high-magnitude mini-clusters-- features not present in the diffuse coefficient spread for C$_{60}$. The likely source of this spread is the multitude of randomized short-range orientations of individual C$_{60}$ atoms (inset of Fig. \ref{fig:Fig7}). We postulate that the diffuse spread leads to complex frequency-domain interactions with sampling patterns and the thresholding function, conditions that make it more difficult to tune the algorithm's parameters. 

% The low-frequency cluster tends to decrease with image dimensions (Fig. \ref{fig:Fig1}), indicating that with increasing dimensions a lower threshold ratio is needed to utilize a large enough number of coefficients for quality reconstruction. This manifests in Fig. \ref{fig:Fig7}, where 

% VERIFY WHAT HAPPENS AT LOW THRESH / WOULD BE GOOD TO SEE HOW HIGH SSIMS AT LOW THRESH FOR C60 ARE ACTUALLY JUST HEAVY LOW PASS FILTERS

% [DCTs of line patterns showed more low-frequency content than for Lissajous; can I connect c60 behavior for sig vs thresh to this trend]
% [Want to say: diffuse spread of coefficients makes it more likely that sampling pattern artifacts will be present in reconstruction]

% Across our simulations, we find that SSIM generally correlates well with visual perception of reconstruction quality, and serves this purpose better than MSE. Indeed, SSIM was constructed to judge the preservation of structure in a reconstructed image, which is a task for which human eyes are naturally well-suited. 

In our studies, SSIM proved to be a faithful reconstruction quality metric in terms of capturing the influence of unwanted artifacts. Reconstructions were also evaluated for MSE, another commonly used quality metric. MSE lacks SSIM's useful universal scale, making cross-comparison of images and phase diagrams more difficult. Furthermore, MSE is not adept at capturing structural artifacts \cite{Wang2009}, and this flaw is displayed in phase diagrams created using the metric (Fig. \ref{fig:MSE_Phase_Diagrams}). While they moderately resemble those for SSIM, these diagrams fail to properly differentiate between good reconstructions and those marred by artifacts. As a particularly harsh example, the poorly reconstructed image of TCNQ at noise and sampling density equal to 0.1 gives a poor SSIM; the MSE, however, is given a median value. A reverse effect occurs for FeSe at these parameters, but visual inspection of the reconstruction yields long-scale structure largely intact, seemingly further confirming SSIM's utility. However, small-scale structure, i.e. the lattice and defects, are perturbed, and MSE may be better for capturing such anomalies.
% The chosen images contain three distinct defect types. The TCNQ monolayer (Fig. \ref{fig:Fig1}b) contains a single molecular defect with length approximately 20\% of the image span, while the FeSe lattice (Fig. \ref{fig:Fig1}a) contains multiple Se vacancies with each around 5\% of the image span. In contrast to these point vacancies, the C$_{60}$ bilayer (Fig. \ref{fig:Fig1}c) features a step running across the image. 
% who developed ideas behind phase transitions
% [discuss difference plots]
% \section{Improvements} \label{sec:improvements}
% % algorithm
% Approximate message passing, related to Bayesian inference, is also used. Other algorithms include LASSO [cite Japan].
% [other algorithms used] [talk about pros/cons]
% [mention other reconstructions, eg gaussian processing?] 
% Different reconstruction metrics: PSNR, etc

\section{Conclusions} \label{sec:discussion}
Our results show that there are significant benefits for using CS for STM, which should also extend to other scanning probe microscopies. Reduction in the acquisition time can be sizeable, allowing for more efficient sampling of materials, with greater extent and higher probability to locate regions of interest. This methodology is readily applicable to imaging of periodic structures, but also to defects and imperfections.  We intentionally used a simple framework to set-up a baseline on which future improvements in CS reconstruction can be made. It is clear that with proper thresholding initialization, satisfactory reconstruction can be obtained without the presence of sampling pattern artifacts. However, in order to properly set the weights, it is advisable to inform the model with prior imaging of a similar sample. However, CS is a highly extensible framework open to more intelligent and in-situ approaches to determine the most effective sampling path and select successful algorithm parameters and transform matrices. Intriguingly, ``anomalies" in CS reconstruction, such as the ones we observed with C$_{60}$ may signal interesting properties of the material, such as the lack of true long-range order or dynamic processes in the experiment, which can then be studied with higher fidelity.

% he algorithm's stopping condition can be related to a bound on the noise variance. 

% Lissajous pattern is not ideal for center defects as not as many points there

% These effects are likely due to the difference in frequency content of the different sampling patterns.

%\bibliographystyle{apsrev4-1-BL-mod}
%\bibliography{references}


\begin{acknowledgments}
We gratefully acknowledge Seokmin Jeon and Simon Kelly for their help with sample preparation for STM experiments with adsorbed molecules. 

Data analysis and interpretation was sponsored by the U. S. Department of Energy, Office of Science, Basic Energy Sciences, Materials Sciences and Engineering Division. Experimental data was acquired at the Center for Nanophase Materials Sciences, which is a DOE Office of Science User Facility. Student (BEL, AFG) research support was provided by the DOE Science Undergraduate Laboratory Internships (SULI) program.
\end{acknowledgments}

% % Appendix
% \appendix*
% \section{Additional figures} \label{sec:appendix}
% % \clearpage

% \begin{figure}
%     \centering
%     \includegraphics[width=\columnwidth]{img/App_Lissajou_SSIM-Phase-Diagrams.png}
%     \caption{Noise perturbation intensity ($\delta$) vs. sampling density ($\rho$) SSIM phase diagrams for reconstructions of TCNQ, C$_{60}$ and FeSe (l-r) for the Lissajous sampling pattern with defect phase diagrams in the inset.}
%     \label{fig:Lissajous_SSIM_Phase_Diagrams}
% \end{figure}
% \begin{figure}
%     \centering
%     \includegraphics[width=\columnwidth]{img/App_Lissajou_Reconstructions.png}
%     \caption{Reconstructed images for ten, five, and two-fold undersampling (l-r) for the Lissajous sampling pattern and 100 iterations, with magnified defects in insets.}
%     \label{fig:Lissajou_Reconstructions}
% \end{figure}

\end{document}